\documentclass[prb,aps,amsmath,twocolumn,superscriptaddress,floatfix]{revtex4-2}
\usepackage{amsmath}
\usepackage{amssymb}
\usepackage{bbm}
\usepackage{dsfont}
\usepackage{graphicx}
\usepackage{color}
\usepackage[svgnames]{xcolor}
\usepackage[normalem]{ulem}
\begin{document}

\title{Negative critical currents in single-channel Josephson junctions}

\author{A. V. Rozhkov}
\affiliation{Institute for Theoretical and Applied Electrodynamics, Russian
Academy of Sciences, 125412 Moscow, Russia}

\author{Tony Liu}

\affiliation{Department of Physics, University of Washington, Seattle, WA
98195 USA}

\author{A. V. Andreev}

\affiliation{Skolkovo Institute of Science  and  Technology,  Moscow,
143026,  Russia}

\affiliation{Department of Physics, University of Washington, Seattle, WA
98195 USA}

\affiliation{L. D. Landau Institute for Theoretical Physics, Moscow, 119334
Russia}

\author{B. Z. Spivak}

\affiliation{Department of Physics, University of Washington, Seattle, WA
98195 USA}

\begin{abstract}
We argue that negative critical currents arise generically in Josephson
junctions formed by single channel conductors. Specifically, we
theoretically study the Josephson coupling between two superconducting
leads connected by a one-dimensional conductor in the Coulomb blockade
regime. We show that in the clean regime the sign of the critical current
alternates with the number of electrons in the normal region. For odd
occupancy the critical current is negative even when the number of
electrons on the conductor is large.
\end{abstract}

\date{\today}

\maketitle

The energy of a superconductor-normal metal-superconductor (SNS) junction
depends on the order parameter phase difference between the two
superconductors. In the limit of weak tunneling between the normal region
and the superconductors, this dependence has the form
\begin{equation}
\label{eq:J_c_def}
E_{\rm J} (\chi_{1}-\chi_{2})
=
- \frac{\hbar J_{c}}{2 e} \cos(\chi_{1}-\chi_{2}).
\end{equation}
Here
$\chi_{1}$
and
$\chi_{2}$
are the order parameter phases in the two superconductors, $e$ is the
electron charge, and $J_c$ is the critical current of the junction.

It is possible to prove that in the single-particle approximation
the critical current is always
positive~\cite{Beenakker}.
Beyond the non-interacting electron approximation, there are no general
principles which determine the sign of
$J_{c}$.
Several physical mechanism of negative currents have been proposed. The
sign of the critical current of a superconductor-ferromagnet-superconductor
junction is an oscillating function of the magnetization and the length of
ferromagnet, see
Ref.~\onlinecite{Buzdin_review,Ryazanov}.
Even in the absence of macroscopic magnetization, critical current can be
negative if it is mediated by tunneling through a magnetic
impurity~\cite{Bulaevskii-Kuzii-Sobyanin},
a resonant
state~\cite{Glazman-Matveev,Kivelson-Spivak,Rozhkov-magnetic},
or a quantum dot in the Coulomb blockade
regime~\cite{Rozhkov}.
Recently, negative critical currents were observed in SNS Josephson
junctions with normal region comprised of carbon
nanotube~\cite{Bouchiat}
and
semiconductors~\cite{Kouwenhoven,Marcus-spinful,Marcus}.

We consider an SNS junction formed by a one-dimensional metallic wire in
the Coulomb blockade regime. We show that alternation of the sign of the
critical current as a function of the number of electrons in the normal
region is a generic property of such systems. Namely, the critical current
is positive when the number of electrons is even, and negative if it is
odd, even when the number of electrons in the junction is large.

The physical reason for the sign alternation can be traced to the node
theorem for electron wave functions in one-dimension (see, for
example~\cite{Landau}).
According to this theorem, the number of nodes in the wave function of an
energy eigenstate is given by the ordinal number of the energy level
counted from the ground state. Indeed, we note that the amplitude of the
electron tunneling through an insulating barrier from a single-particle
state in the normal region with wave function 
$\psi_m(x)$
to a state in the lead 
$i = 1,2$
with wave function 
$\phi^{(i)}_k(x)$
may be expressed
as~\cite{Landau,Bardeen}
\begin{equation}
\label{eq:t_km}
 t^{(i)}_{mk} = \frac{1}{2 m^*}\!\!\left.
		 \left[
			\phi^{(i)}_k(x) \partial_x \psi_{m}(x)
			-
			\psi_{m}(x) \partial_x \phi^{(i)}_k(x)
		\right] \right|_{x=x_{{i}}}.
\end{equation}
Here
$m^*$
denotes the electron mass, the $x$-axis is along the wire, and
$x_{{i}}$
is located inside the tunneling barrier between the wire and lead $i$;
$x_{{1}} =0$,
and
$x_{{2}} = L$,
where $L$ is the length of the wire. We assume that all electron wave
functions can be chosen real. In second order perturbation theory the
single particle tunneling amplitude between the two leads through the
virtual state $m$ in the wire is proportional to the product of derivatives
$\partial_x \psi_m(x)$
at the contacts,
\begin{eqnarray}
\label{eq:sign_alternation}
[\partial_x \psi_m (0)][\partial_x \psi_m (L)]
\!=\!
(-1)^{m+1} | \partial_x \psi_m (0)||\partial_x \psi_m (L)|.
\quad
\end{eqnarray}
Alternation of the sign in this equation is a direct consequence of the
node theorem.

To elucidate the mechanism of the critical current sign alternation we
consider a system described by the following Hamiltonian
\begin{eqnarray}
\label{eq::Hbare}
\hat{H}
=
E_{C} (\hat{N}-{\cal N}_0)^2
+ \hat{H}^{\rm 1D}
+\sum_{i=1,2}\left[\hat{H}^{(i)}_{t}
+\hat{H}^{(i)}_{\rm SC}\right].
\end{eqnarray}
In this expression,
$\hat{H}^{(i)}_{\rm SC}$
is the Hamiltonian of the $i$'th superconducting lead,
$\hat{H}^{\rm 1D}$
is the Hamiltonian of the normal metal region
\begin{eqnarray}
\hat{H}^{\rm 1D}
=
\sum_{m,\sigma}
	\xi_{m}^{\vphantom{\dagger}}
	c^\dag_{\sigma m} c^{\vphantom{\dagger}}_{\sigma m}.
\end{eqnarray}
Here
$c_{\sigma m}$
is the annihilation operator of an electron with spin
$\sigma=\uparrow,\downarrow$
and the $m$-th single-electron state, whose energy
$\xi_{m}^{\vphantom{\dagger}}$
is measured relative to the chemical potential (we will assume that
$\xi_m$
is a monotonic function of index $m$). The mean level spacing for the
conductor is
$\delta \sim \hbar v_{\rm F}/L$,
where
$v_{\rm F}$
is the Fermi velocity.

The tunneling Hamiltonian may be expressed in terms of the tunneling matrix
elements in
Eq.~\eqref{eq:t_km}
as
\begin{equation}
\hat{H}_{t}^{(i)}
=
\sum_{m,k, \sigma } t^{(i)}_{mk} c^\dagger_{\sigma m} a^{(i)}_{\sigma k} +
{\rm H. c.},
\end{equation}
where
$a^{(i)}_{\sigma k}$
denotes the electron annihilation operator in state $k$ in the
superconducting lead $i$.

To keep the presentation more transparent we treat the electron
interactions in the ``zero mode" approximation. In doing so we neglect the
correlations induced by the electron-electron interactions inside the
normal region. This approximation is applicable for small ratio between
electron potential and kinetic energy:
$r_{s}=e^2/\hbar v_F\ll 1$.
Furthermore, we assume that the length of the metallic wire $L$ satisfies
the inequality,
$r_s \ln (k_F L) \ll 1$
so that the Tomonaga-Luttinger liquid
effects~\cite{Kane_Fisher,Matveev_tunneling_TL,Kane_Fisher_Balents}
may be neglected.

The ``zero mode" interaction is represented by the first term in
Eq.~\eqref{eq::Hbare}.
Here
$E_{C}\sim e^{2}/L \ll \delta$
is a single-electron charging energy, operator
$\hat{N}=\sum_{\sigma m}c^\dag_{\sigma m} c^{\vphantom{\dagger}}_{\sigma m}$
counts the number of electrons in the 1D conductor. (In the absence of
tunneling this number is quantized.) The parameter
${\cal N}_0$
is proportional to the gate voltage and controls the number of electrons in
the wire. We work in the approximation of weak tunneling and assume that
the system is sufficiently far from the charge degeneracy point, so that
quantum charge fluctuations are small. In this case the spin in the ground
state is $0$ for an even number of electrons and 1/2 for an odd number.

The lowest order of perturbation theory with respect to
$H^{(i)}_{t}$
which yields the dependence of the junction energy on
$(\chi_{1}-\chi_{2})$
is fourth. For simplicity we assume that $\Delta$, the quasiparticle energy
gap in superconductors, exceeds both the Coulomb energy and the mean level
spacing:
$\Delta \gg E_{C},\,\delta$.
In this regime quasiparticles can tunnel from superconductors to the normal
metal wire only by pairs, and the part of the pair-tunneling Hamiltonian
between superconductor $i$ and the metallic wire may be written in the form
\begin{eqnarray}
\label{eq::H_Ti}
\hat{H}^{(i)}_{\rm T} =
e^{i\chi_i}
\sum_{mn}
	T^{(i)}_{mn} c^\dag_{\uparrow m} c^\dag_{ \downarrow n}
+ {\rm H.c.},
\end{eqnarray}
where
$T^{(i)}_{mn}$
denotes the tunneling amplitude of a Cooper pair from lead $i$ into the
states $m$ and $n$ in the normal region. For low-lying excited states $m$
and $n$ in the wire, satisfying
$\Delta \gg |\xi_n|, |\xi_m|$,
amplitude
$T^{(i)}_{mn}$
can be expressed in terms of the single-particle tunneling amplitudes in
Eq.~\eqref{eq:t_km}
in the
form~\cite{Nazarov-Hekking,Wilkins}
\begin{equation}
\label{eq::tunn_ampl}
 T^{(i)}_{mn}
=-
\sum_k
	\frac{ t_{km}^{(i)} t_{kn}^{(i)}\,
	|\langle a_{\uparrow k} a_{\downarrow k} \rangle |}{\epsilon_k}.
\end{equation}
In this approximate expression,
$\epsilon_k$
is the quasiparticle energy in state $k$ of the superconductor, and
$\langle a_{\uparrow k} a_{\downarrow k}\rangle$
denotes the Cooper pair condensation amplitude. Performing the summation
over $k$ in
Eq.~(\ref{eq::tunn_ampl}),
one obtains the following estimate
\begin{eqnarray}
\label{eq::Lambda}
|T^{(i)}_{mn}|
\propto
g^{(i)} \delta, 
\end{eqnarray}
where
$g^{(i)}$
is dimensionless conductance of the $i$-th SN interface.

Once higher-energy degrees of freedom are ``integrated out", the effective
model Hamiltonian reads
\begin{eqnarray}
\label{eq::H1D}
\hat{H}_{\rm eff}
=
E_{C} (\hat{N}-{\cal N}_0)^2 + \hat{H}^{\rm 1D}
+
\hat{H}^{(1)}_{\rm T }+ \hat{H}^{(2)}_{\rm T}.
\end{eqnarray}
Note that it only contains degrees of freedom in the normal metal.


We evaluate the Josephson coupling energy of the system using second-order
perturbation theory in powers of
$\hat{H}_{\rm T}^{(1,2)}$
\begin{eqnarray}
\label{eq::pert_theor}
E_{J}\!
=
\!
\sum_{mn}
	\frac{ \langle {{{0}}}|
		\hat{H}^{(1)}_{\rm T}
		|\!\uparrow \! m, \downarrow\! n\rangle
	\langle \uparrow \! m,\downarrow \!n|
		\hat{H}^{(2)}_{\rm T}
	|{\rm {{0}}}\rangle}{E_{\rm {{0}}} - E_{mn}}
+ {\rm c.c.},
\quad
\end{eqnarray}
where
$|{\rm {{0}}}\rangle$
is the ground state of the wire, while
$|\!\!\uparrow \!m,\downarrow \!\! n \rangle$
is the two-particle excited state characterized by the presence (or
absence) of two electrons with opposite spins, one on level $n$, another on
level $m$. The structure of the ground state
$|{\rm {{0}}}\rangle$
depends on the parity of
$N_0$.
When number of electrons is even,
$N_0 = 2M$,
the ground state
$\left| {\rm {{0}}} \right>$
is spin singlet, and any
$m \geq M$
level is empty, while  any
$0 < m < M$
level is doubly occupied. This ground state remains stable as long as the
gate potential
${\cal N}_0$
satisfies the following inequalities
\begin{eqnarray}
\label{eq::even_stability}
- E_{C} - \xi_M < 2 E_{C} ( 2M - {\cal N}_0 ) < E_{C} - \xi_{M-1}.
\end{eqnarray}
Alternatively, for a fixed gate potential
${\cal N}_0$,
one can view this relation as a condition on
$N_0 = 2M$.

When either of strict inequalities in
Eq.~(\ref{eq::even_stability})
become an equality, the ground state becomes charge-degenerate. For
example, if
$2 E_{C} ( 2M - {\cal N}_0 ) = - E_{C} - \xi_M$,
the state with
$2M$
electrons and a state with
$2M + 1$
electrons become degenerate. The extra electron occupies
$m=M$
level. The ground state with odd
$N_0 = 2M + 1$
is stable when
\begin{eqnarray}
|2 E_{C} (2M + 1 - {\cal N}_0) + \xi_M| < E_{C}.
\end{eqnarray}
For odd
$N_0$,
the ground state is spin doublet.

In the following, we will re-define single-electron index $m$ as follows
$m \rightarrow m-M$.
That way, all levels with
$m<0$
are doubly occupied, all levels with
$m>0$
are empty. A single electron resides on
$m=0$
level for odd
$N_0$,
otherwise, this level is empty. Furthermore, without loss of generality, we
can assume that
$\xi_0 = 0$.

Ultimately, the perturbation-theory
expression~(\ref{eq::pert_theor})
for the Josephson coupling
$E_{\rm J}$
can be written as
\begin{eqnarray}
\label{eq::EJ_general}
E_{\rm J} = - ( E_{+} + E_{-} ) \cos (\chi_1 - \chi_2),
\end{eqnarray}
where energies
$E_{\pm}$
represent two-electron and two-hole contributions
\begin{eqnarray}
\label{eq::E_even_pm_fin}
E_{\pm}
=
\sum_{mn}
	\frac{2 T^{(1)}_{mn} T^{(2)}_{mn}
		\Theta(\pm m \pm 1/2) \Theta (\pm n \pm P/2)}
	{4E_{C} \pm [4 E_{C} (N_0 - {\cal N}_0) + \xi_{m} + \xi_{n}]},
\end{eqnarray}
where the choice of the sign (top/bottom) in the right-hand side is
dictated by the sign in the left, and the half-integer terms in the
arguments of the Heaviside functions
$\Theta(x)$
are introduced to avoid the uncertainty of
$\Theta(0)$.
Note that in
Eq.~(\ref{eq::E_even_pm_fin})
the summation range for index $n$ depends on the parity $P$ defined as
\begin{eqnarray}
\label{eq::P_def}
P= (-1)^{N_0}.
\end{eqnarray}
For odd
$N_0$,
this accounts for the Pauli blocking of
$n=0$
single-electron state by a single electron occupying this state.

Another crucial observation about the sum in
Eq.~\eqref{eq::E_even_pm_fin}
is the sign alternation of the terms being summed. While the denominators
are always positive, the numerators signs demonstrate a different pattern
\begin{equation}
\label{eq:T_sign}
T^{(1)}_{mn} T^{(2)}_{mn}
=
(-1)^{m+n} |T^{(1)}_{mn} T^{(2)}_{mn}|.
\end{equation}
To justify the sign-alternating factor
$(-1)^{m+n}$
in this relation one can use
Eqs.~(\ref{eq:t_km})
and~(\ref{eq::tunn_ampl})
to express
$T^{(i)}_{mn}$
in terms of the derivatives
$\partial_x \psi_m|_{x=0,L}$
and
$\partial_x \psi_n|_{x=0,L}$,
and then apply the node
theorem~(\ref{eq:sign_alternation})
to the product
$T^{(1)}_{mn} T^{(2)}_{mn}$.


So far we did not make any assumptions about the strength of disorder in
the wire. Now we apply the formalism to the clean-wire case,
$L\gg l$,
where $l$ is the elastic electron mean free path. In such a situation, the
wave functions are
$\psi_m \propto \sin (k_m x)$,
where the quantized momentum is
$k_m = k_{\rm F} + \pi m /L$,
and the Fermi momentum equals to
$k_{\rm F} \approx \pi N_0/(2L)$
for
$N_0 \gg 1$.
The single-particle energies are
$\xi_m = m \delta$.

In this regime,
$|T^{(1)}_{mn} T^{(2)}_{mn}|$
may be considered independent of $m$ and $n$, while the sign of the
product
$T^{(1)}_{mn} T^{(2)}_{mn}$
satisfies
Eq.~(\ref{eq:T_sign}).
Thus,
Eq.~(\ref{eq::E_even_pm_fin})
for even number of electrons can be expressed as
\begin{eqnarray}
\label{eq::clean_Josephson_coupling}
E^{\rm (even)}_{\pm}
=
E_0 \sum_{m \geq 0 \atop n \geq 0}
	\frac{(-1)^{m+n}} {\kappa_\pm + m + n}.
\end{eqnarray}
Here
$E_0 \propto g^{(1)} g^{(2)} \delta$,
and the dimensionless offset parameters are
\begin{eqnarray}
\label{eq::offset_kappa}
\kappa_{\pm} = \frac{4E_{C}}{\delta}
	\left[
		1 \pm (N_0 - {\cal N}_0) + \frac{(1\mp 1)\delta }{4 E_{C}}
	\right].
\end{eqnarray}
Note that
$\kappa_\pm > 0$,
as ensured by
inequalities~(\ref{eq::even_stability}),
and
$\kappa_{+} + \kappa_{-} \approx 2$.
The latter relation means that at least one of $\kappa$'s is of order
unity, and neither of them exceed 2.

Therefore, we reduce the issue of finding the Josephson coupling to the task
of evaluating the sum in
Eq.~(\ref{eq::clean_Josephson_coupling}).
To proceed, we rewrite this double sum in the form
\begin{eqnarray}
\label{eq::sum_of_f}
{\cal S}_{\pm} = \sum_{n = 0}^{+\infty} (-1)^n f (\kappa_\pm + n),
\\
\label{eq::f_def}
\text{where}\quad
f (y) = \sum_{m = 0}^{+\infty} \frac{(-1)^{m}} {y + m}.
\end{eqnarray}
Since the right-hand side of
Eq.~(\ref{eq::f_def})
is a sign-alternating series satisfying the Leibniz criterion [sequence
$(y + m)^{-1}$
monotonically decreases to zero for growing index $m$], we conclude that,
for positive $y$, the series is convergent, and function
$f(y)$
is finite. Moreover, the Leibniz theorem guarantees that, for positive $y$,
$f(y) > 0$
since the first term in the
sum~(\ref{eq::f_def})
is positive. Additionally, it is easy to prove that
$f(y)$
decreases monotonically when
$y \rightarrow +\infty$.
Indeed, the derivative of $f$
\begin{eqnarray}
\label{eq::f_derivative}
f' (y) =
\sum_{\ell = 0}^{+\infty}\left[
	\frac{1}{(y + 2 \ell + 1)^2} - \frac{1}{(y + 2 \ell)^2}\right]
\end{eqnarray}
is negative since it is a convergent series of strictly negative terms.
Therefore, the series in
Eq.~(\ref{eq::sum_of_f})
also passes the Leibniz test. Furthermore,
${\cal S}_\pm > 0$
since
$f (\kappa_\pm)$
are both positive. Thus we conclude that for an even number of electrons in
the wire the critical current
$J_c$
in
Eq.~\eqref{eq:J_c_def}
is positive.

Let us now consider the situation with an odd number of electrons in the
wire. In this case, using
Eq.~\eqref{eq::E_even_pm_fin}
with
$P=-1$,
we obtain the following expression for
$E_\pm$
[see
Eq.~\eqref{eq::EJ_general}]
\begin{eqnarray}
\label{eq::clean_Josephson_coupling_odd}
E^{\rm (odd)}_{\pm}
=
- E_0 \sum_{m \geq 0 \atop n \geq 0}
	\frac{(-1)^{m+n}} {(\kappa_\pm \pm 1) + m + n}.
\end{eqnarray}
This relation is the odd-$N_0$ counterpart of
Eq.~(\ref{eq::clean_Josephson_coupling}).
The differences between these two expressions, both in terms of the overall
sign and the denominator structure, follows from the difference in values
of $P$ for even/odd
$N_0$.

The argumentation presented above for the even-$N_0$ case is trivially
applicable for
Eq.~(\ref{eq::clean_Josephson_coupling_odd}).
This allows us to conclude that
$E_{\pm}^{\rm (odd)} < 0$.
Thus, for an odd number of electrons in the wire the critical current
$J_c$
is negative. In other words, an addition or subtraction of a single
electron from the conductor changes the sign of the critical current even
in the case where the number of electrons is large.

The reason for the opposite sign of the critical current
$J_c$
in the cases of even and odd number of electrons in the wire
$N_0$
is related to the fact that the signs of the corresponding alternating
series,
Eqs.~(\ref{eq::clean_Josephson_coupling})
and~(\ref{eq::clean_Josephson_coupling_odd}),
are determined by the signs of the terms with the smallest energy
denominator.

So far we considered the case of a clean metallic wire. In the presence of
a disorder potential
$V(x)$,
the general perturbative
expression~(\ref{eq::E_even_pm_fin})
remains valid.
Equation~\eqref{eq:T_sign},
which follows from the node theorem, is valid as well. However, generally
speaking, the absolute values of the terms in the alternating series
Eq.~(\ref{eq::E_even_pm_fin})
do not decrease monotonically since the products
$|T^{(1)}_{mn} T^{(2)}_{mn}|$
become sample-specific functions of $n$ and $m$. Therefore, the double sum
in
Eq.~(\ref{eq::E_even_pm_fin})
may be dominated by the states whose wave functions have strongest coupling
to the leads, rather then those with smallest energy denominators. In this
case the value of the critical current depends on the realization of
disorder.

Interestingly, it is possible to prove that, similarly to the case of
non-interacting
electrons~\cite{Beenakker},
the Josephson coupling for an even number of electrons in the wire remains
positive for any disorder in the wire. Our proof relies on factorization
$T^{(1)}_{mn} T^{(2)}_{mn} = s_n s_m$,
where function
$s_n \propto \partial_x \psi_n (0) \partial_x \psi_n (L)$.
We also make use of the following integral representation of a fraction
$1/x = \int_0^{+\infty} du {\rm e}^{-ux}$
valid for any
$x>0$.
Applying the latter representation to the denominator in
expression~(\ref{eq::E_even_pm_fin})
one derives
\begin{eqnarray}
\label{eq::Epm_integ}
E_{\pm}^{\rm (even)}
=
\int_0^{+\infty} \!\!du\,
	G_{\pm}^2 {\rm e}^{ - 4E_{C} [1 \pm (N_0 - {\cal N}_0)]u },
\end{eqnarray}
where functions
$G_{\pm} = G_{\pm} (u)$
are defined for
$u>0$
by convergent series
\begin{eqnarray}
\label{eq::Ge}
G_{\pm} = \sum_{m}
	 (-1)^m {\rm e}^{ \mp u \xi_m} |s_m| \Theta(\pm m \pm 1/2).
\end{eqnarray}
It is clear from
Eq.~(\ref{eq::Epm_integ})
that
$E_{\pm}^{\rm (even)}$
are both positive for any spectrum and the wave function structure.

Finally, we would like to make the following observations.

(i)~In the case where the number of electrons in the wire is odd and in the
presence of disordered potential the critical current of the junction has
random sign, and the probability of the negative sign decreases as the
strength of the disorder in the wire increases. For example, in the
strongly disordered case where the electron mean free path $l$ is shorter
than the wire length,
$l < L$,
the electron wave functions are localized. In this case  the distribution of
$|T^{(1)}_{mn} T^{(2)}_{mn}|$
in
Eq.~\eqref{eq:T_sign}
is exponentially broad. As a result, the double sum for the critical
current in
Eq.~\eqref{eq::E_even_pm_fin}
is dominated by contributions of states whose wave functions have strongest
coupling to both leads. Since such states are typically either doubly
occupied or empty, the probability of negative critical current is small.

(ii)~Since the results obtained above are based on the one-dimensional
node theorem, they are of purely one-dimensional character. In general, in
the systems where
$E_{F}\gg \delta$
and in the case where the normal metal stripe has a finite width larger
than the Fermi wave length, the probability for the critical current to be
negative decreases when the sample width grows.

(iii)~For longer conductors,
$L\gg L_{TL}$,
the Luttinger liquid effects become significant. In this regime the
0-junction critical current
$J_{c}(L)$
has been extensively studied
theoretically~\cite{Maslov,Aflek,Takane}.
In this limit the Luttinger liquid effects change the $L$ dependence of the
critical current from the single particle dependence $1/L$ to a power of
$L$ which depends on the value of interaction constant. As far as a
question of the sign of
$J_{c}$ is concerned, we conjecture that the Luttinger liquid effects do not
destroy the alternation of the Josephson coupling sign.   In particular, at  $r_{s}\gg 1$, where the system is close to an antiferromagnetic Wigner crystal, an extension of the  arguments  presented in Ref.~\onlinecite{Kivelson-Spivak}  suggests that the sign of the critical current should  oscillate as a function of the number of electrons.  We also would like to mention that the sign oscillation of the pair field correlator as a function of distance in 1D Kondo system has been discussed in Ref.~\onlinecite{Berg}.

To summarize, we showed that in clean single channel Josephson junctions
the sign of the critical current alternates with the number of electrons in
the normal region, being positive if the  number of electrons is even  and
negative when it is odd.

\begin{acknowledgments}
B.Z.S. acknowledges useful discussions with S. Kivelson.
The work of T.L. and A.V.A. was supported by the National Science
Foundation Grant MRSEC DMR-1719797.

\end{acknowledgments}


\begin{thebibliography}{99}

\bibitem{Beenakker} M. Titov, Ph. Jacquod, and C.W.J.~Beenakker,
\emph{Negative superfluid density: Mesoscopic fluctuations and reverse of
the supercurrent through a disordered Josephson junction}, Phys. Rev. B
{\bf 65}, 012504 (2001).

\bibitem{Ryazanov} V.V. Ryazanov, V.A. Oboznov, A.Yu.~Rusanov,
A.V.~Veretennikov, A.A.~Golubov, and J.~Aarts, \emph{ Coupling of Two
Superconductors through a Ferromagnet: Evidence for a $\pi$ Junction},
Phys. Rev. Lett. {\bf 86} 2427, (2001).

\bibitem{Buzdin_review} A.I.~Buzdin, \emph{Proximity effects in
superconductor-ferromagnet heterostructures}, Rev. Mod. Phys. \textbf{77},
935 (2005).

\bibitem{Kouwenhoven} J.A.~Van Dam, Y.V.~Nazarov, E.P.A.M.~Bakkers, S.~De
Franceschi, and L.P.~Kouwenhoven, \emph{Supercurrent reversal in quantum
dots},  Nature (London) \textbf{442}, 667 (2006).


\bibitem{Bouchiat} J.P.~Cleuziou, W.~Wernsdorfer, V.~Bouchiat,
T.~Ondar\c{c}uhu, and M.~Monthioux, \emph{Carbon nanotube superconducting
quantum interference device}, Nat. Nanotechnol. {\bf 1}, 53 (2006).

\bibitem{Marcus-spinful} W. Chang, V.E.~Manucharyan, T.S.~Jespersen,
J.~Nyg\o{a}rd, and C.M.~Marcus,   \emph{Tunneling Spectroscopy of
Quasiparticle Bound States in a Spinful Josephson Junction}, Phys. Rev.
Lett. \textbf{110}, 217005 (2013).

\bibitem{Marcus} D. Razmadze, E.C.T.~O'Farrell, P.~Krogstrup, and
C.M.~Marcus, \emph{Quantum Dot Parity Effects in Trivial and Topological
Josephson Junctions}, Phys. Rev. Lett. \textbf{125}, 116803 (2020).

\bibitem{Landau} L.D. Landau, and E.M. Lifshitz, \emph{Quantum Mechanics
Non-Relativistic Theory. Course of Theoretical Physics}, Third Edition,
Volume 3, (Butterworth-Heinemann, Oxford, 1981).

\bibitem{Maslov} Dmitrii L. Maslov, Michael Stone, Paul M.~Goldbart, and
Daniel Loss, \emph{Josephson current and proximity effect in Luttinger
liquids}, Phys. Rev. B {\bf 53}, 1548 (1996)

\bibitem{Bardeen} J. Bardeen, \emph{Tunnelling from a Many-Particle Point
of View}, Phys. Rev. Lett. \textbf{6}, 57 (1961).

\bibitem{Aflek} Domenico Giuliano and Ian Affleck, \emph{The Josephson
current through a long quantum wire}, J. Stat. Mech. (2013) P02034.

\bibitem{Takane} Y. Takane, \emph{DC Josephson Effect in a
Tomonaga–Luttinger Liquid}, J. Phys. Soc. Jpn. {\bf 71}, 550 (2002).

\bibitem{Bulaevskii-Kuzii-Sobyanin} L.N. Bulaevskii, V.V. Kuzii,
A.A.~Sobyanin, \emph{Superconducting system with weak coupling to the
current in the ground state}, JETP Lett. \textbf{25}, 290 (1977);
\emph{On possibility of the spontaneous magnetic flux in a Josephson
junction containing magnetic impurities}, Solid State Commun.  \textbf{25},
1053 (1978).

\bibitem{Glazman-Matveev} L.I. Glazman, and   K.A. Matveev, \emph{Resonant
Josephson current through Kondo impurities in a tunnel barrier}, JETP
Lett. \textbf{49},  659 (1989).

\bibitem{Kivelson-Spivak} B. I. Spivak and S. A. Kivelson, \emph{Negative
local superfluid densities: The difference between dirty superconductors
and dirty Bose liquids}, Phys. Rev. B {\bf 43}, 3740(R) (1991).

\bibitem{Rozhkov-magnetic}  A.V. Rozhkov and Daniel P.~Arovas,
\emph{Josephson Coupling through a Magnetic Impurity}, Phys. Rev. Lett.
\textbf{82}, 2788 (1999).

\bibitem{Rozhkov}  A.V. Rozhkov, Daniel P.~Arovas, and F.~Guinea,
\emph{Josephson coupling through a quantum dot},  Phys. Rev. B \textbf{64},
233301 (2001).

\bibitem{Kane_Fisher} C.L. Kane and M.P.A.~Fisher, \emph{Transport in a
one-channel Luttinger liquid}, Phys. Rev. Lett. \textbf{68}, 1220 (1992).

\bibitem{Matveev_tunneling_TL} K.A. Matveev, Dongxiao Yue, and L.I.~Glazman,
\emph{Tunneling in one-dimensional non-Luttinger electron liquid}, Phys.
Rev. Lett. \textbf{71}, 3351 (1993).

\bibitem{Kane_Fisher_Balents} C. L.  Kane, Leon Balents, and
Matthew P.A.~Fisher, \emph{Coulomb Interactions and Mesoscopic Effects in
Carbon Nanotubes}, Phys. Rev. Lett. \textbf{79}, 5086 (1997).

\bibitem{Nazarov-Hekking} F.W.J. Hekking and Yu.V. Nazarov, \emph{Subgap
conductivity of a superconductor–normal-metal tunnel interface},
Phys. Rev. B \textbf{49}, 6847 (1994).

\bibitem{Wilkins} J.W. Wilkins, \emph{Multiparticle Tunneling}, in
\emph{Tunneling Phenomena in Solids}, edited by E.~Burstein and
S.~Lundqvist (Plenum Press New York, 1969), p.~333.


\bibitem{Berg} Erez Berg, Eduardo Fradkin, and Steven A. Kivelson,  
\emph{Pair-Density-Wave Correlations in the Kondo-Heisenberg Model},
Phys. Rev. Lett.  \textbf{105}, 146403 (2010).


\end{thebibliography}
\end{document}